\documentclass[pss]{wiley2sp} 
\usepackage{amsmath}
\usepackage{bm}              
\usepackage{w-greek}

\providecommand{\WileyBibTextsc}{}
\let\textsc\WileyBibTextsc
\providecommand{\othercit}{}
\providecommand{\jr}[1]{#1}

\begin{document}
\title{Quantum phase transitions in heavy fermion metals and Kondo insulators}
\titlerunning{}
\author{%
 Qimiao Si\textsuperscript{\Ast,\textsf{\bfseries 1}} and
 Silke Paschen\textsuperscript{\Ast,\textsf{\bfseries 2}}}
\authorrunning{Q.\ Si and S.\ Paschen}
\mail{e-mail
 \textsf{qmsi@rice.edu};
 \textsf{paschen@ifp.tuwien.ac.at}
 }
\institute{%
 \textsuperscript{1} Department of Physics and Astronomy, Rice University, Houston, Texas 77005, USA\\
 \textsuperscript{2}\,Institute of Solid State Physics, Vienna University of
Technology, Wiedner~Hauptstr.~8-10, 1040~Vienna, Austria}
\hyphenation{pa-ra-mag-ne-tic par-ti-cu-lar-ly}


\abstract{%
\abstcol{%
Strongly correlated electron systems at the border of magnetism  are of active
current interest, particularly because the accompanying quantum criticality
provides a route towards both strange-metal non-Fermi liquid behavior and
unconventional superconductivity. Among the many important questions is whether
the magnetism  acts simply as a source of fluctuations in the textbook Landau
framework, or instead  serves as a proxy for some unexpected new physics. We put
into this general context the recent developments on quantum phase transitions
in antiferromagnetic heavy fermion metals. Among these are the extensive recent
theoretical and experimental studies on the physics of Kondo destruction in a
class}{of beyond-Landau quantum critical points. Also discussed are the
theoretical basis for a global phase diagram of antiferromagnetic heavy fermion
metals, and the recent surge of  materials suitable for studying
this phase diagram. Furthermore, we  address the generalization of this  global
phase diagram to the case of Kondo insulators, and consider the future prospect
to study the interplay among Kondo coherence, magnetism and topological states.
Finally, we touch upon related issues beyond the antiferromagnetic settings,
arising in  mixed valent, ferromagnetic, quadrupolar, or spin glass
$f$-electron systems, as well as some general issues on emergent  phases near
quantum critical points.
}}
\maketitle


\section{Introduction}\label{Intro}
Electron correlations give rise to a variety of novel phenomena, including
unconventional superconductivity and non-Fermi liquid behavior. Pertinent
materials include cuprate and iron-based superconductors, heavy fermion metals
and organic charge-transfer salts.  In the presence of correlations, we face the
central question of  how the electrons are organized and, in particular, whether
there are principles that are universal among the various classes of the
strongly correlated materials. One such principle, which has come  to the
forefront in recent years, is quantum criticality
\cite{Natphys-qpt08,JLTP-issue10}.

A quantum critical point (QCP) arises at a second-order transition between two
ground  states. The distinct ground states themselves occur due to competing
interactions of a many-body system. The collective fluctuations of the QCP
control the physics of the  quantum critical regime, which corresponds to  a
wide parameter range at non-zero temperature. While quantum criticality is of
interest in its own right, and is being discussed  in a variety of contexts
ranging from insulating magnets to designer materials such  as quantum dots and
cold atom systems, it plays a special role in strongly correlated electron
systems. Quantum criticality provides a mechanism for  
both
the non-Fermi liquid
behavior
and emergent phases such as unconventional superconductivity.

In 
this 
paper, we focus primarily on the issues arising in  strongly
correlated $f$-electron systems. These systems have been serving as a prototype
setting to study the physics of quantum criticality. We will in particular
highlight a class of QCPs in the antiferromagnetic (AF) Kondo lattice systems,
which features the physics of Kondo destruction. The latter also gives rise to a
global phase diagram, involving zero-temperature states  that are distinct not
only by the AF order but also by the nature of the Fermi surfaces. Experimental
probes of such
``beyond-Landau"
QCPs and emergent phases are discussed. We also
consider the analogous issues in Kondo insulators, extend the notion of Kondo
destruction to mixed valent or ferromagnetic settings, and briefly discuss 
quantum criticality in quadrupolar and spin glass  cases. Finally, we address
several issues  concerned with emergent phases around QCPs.


\section{Local moments and competing interactions}\label{KondoLattice}
Heavy fermion systems represent canonical settings in which localized magnetic
moments and  itinerant conduction electrons coexist
\cite{Si-Science10,Loe07.1,Ste01.1,Hewson}.
There are interactions not only among the
local moments but also between the local moments and conduction electrons. The
competition between these interactions gives rise to distinct ground states.
Therefore, heavy fermion systems provide a fertile setting for quantum phase 
transitions.

The Kondo lattice Hamiltonian reads:
\begin{eqnarray}
H_{\rm KL}&=&
 \sum_{ ij } t_{ij}
c^{\dagger}_{i\sigma}
c^{\phantom\dagger}_{j\sigma} \nonumber\\
&&+
 \sum_{ ij } I_{ij}
{\bf S}_i \cdot {\bf S}_j
+  \sum_{i} J_K {\bf S}_i \cdot c^{\dagger}_{i}
\frac{\vec{\sigma}}{2} c^{\phantom\dagger}_{i} \quad .
\label{kondo-lattice-model}
\end{eqnarray}
It describes materials which contain spin-$\frac{1}{2}$ local moments
($\bf{S}$$_i$ at every site $i$) that are physically associated with the
localized $f$ electrons, and a band of conduction electrons ($c_{i\sigma}$). The
parameters $t_{ij}$ and $I_{ij}$ are respectively the  hopping matrix of the
conduction electrons (with bandwidth $W$) and the RKKY exchange interaction
among the local moments (with a characteristic strength $I$), while $J_K$ is the
AF Kondo exchange interaction. We consider the simplest case, corresponding to
one local moment per unit cell, and a filling of  $x$ conduction electrons
per unit cell.

An effective single-ion Kondo energy scale $T_0$ signifies the onset of initial
Kondo screening:
the local moments are essentially decoupled from the conduction electrons
at $T \gg T_0$, but develop dynamical singlet correlations with the conduction electrons
as  $T$ is lowered towards $T_0$.
This provides a means to check the validity of the Kondo lattice
Hamiltonian through measurements at temperatures above or around $T_0$.
Examples are the observation of Curie-Weiss behavior in the magnetic susceptibility,
and a single-ion Kondo peak in the photoemission
\cite{Klein-prl08}
and STM spectra \cite{Ernst-nature11}.

As temperature is further lowered, the system may evolve towards different ground states.
In one ground state, the local moments form a spin singlet with the conduction electrons.
This Kondo entanglement turns each local moment into an
electronic Kondo resonance excitation. 
The excitations occur at low energies,
near the bare Fermi energy of the conduction electrons. These
resonance excitations hybridize with the conduction electrons, forming 
renormalized quasiparticle bands that are separated by a hybridization gap
\cite{Hewson}.

For the generic case of incommensurate fillings with $x<1$, the total electron
count is $1+x<2$. Therefore, the Fermi energy will be in the lower hybridized
band, located in the peak part of the DOS. This gives rise to a heavy Fermi
liquid (FL) state with enhanced quasiparticle mass and a large Fermi surface. 

In the special case of
commensurate filling, $x=1$, the total electron count
is $2$ per unit cell, corresponding to half-filling. The renormalized Fermi
energy must be in the middle of the hybridization gap, leaving the lower
hybridized band completely occupied and the upper one completely empty. The
result is a Kondo insulator
\cite{Aep92.1,Ris00.1,Tsu97.1}.

Ground states other than the Kondo entangled state are also possible.
The local moments are coupled to each other through an
 exchange interaction.
In Eq.\,(\ref{kondo-lattice-model}), this is the RKKY interaction $I$;
up to Sect.\,\ref{Other}, 
we will consider it to be AF.
This exchange interaction promotes AF order. 
Correspondingly, 
the
Kondo and RKKY
interactions compete  against each other
\cite{Doniach,Varma76}.
This
competition is captured by the tuning  parameter $\delta \equiv T_K^0 / I$,
where  $T_K^0 \approx \rho_0^{-1} \exp(-1/\rho_0J_K) $, with $\rho_0$ being the
density of states of the conduction electrons at the Fermi energy, parameterizes
the Kondo interaction.

In experiments, this tuning is accomplished by the application of external
parameters. Typical examples are applying pressure, chemical
substitution or magnetic
field. Applying pressure increases/decreases  $\delta$ in Ce/Yb-based heavy
fermion compounds. Isoelectronic
substitution of the ligands with smaller/larger atoms
has the effect of applying positive/negative chemical pressure, although one
 needs to keep in mind that
substitutions also introduce
 additional disorder.
Finally, an external magnetic field influences the tendencies towards Kondo
screening and AF order to different degrees
(small magnetic fields reduce
the Kondo scale in quadratic order but
are coupled to the AF order linearly), thereby modifying the ratio $\delta$.

The explicit observation of AF QCPs in several heavy fermion materials provides
a setting to study quantum criticality in general. Furthermore, here, quantum
criticality becomes intertwined with two of the key notions of correlated
electrons: non-Fermi liquid behavior and, in many cases, unconventional
superconductivity,
occur in the quantum critical regime of heavy
fermion metals.


\begin{figure}[t!]                                                                
\includegraphics[width=0.9\linewidth]{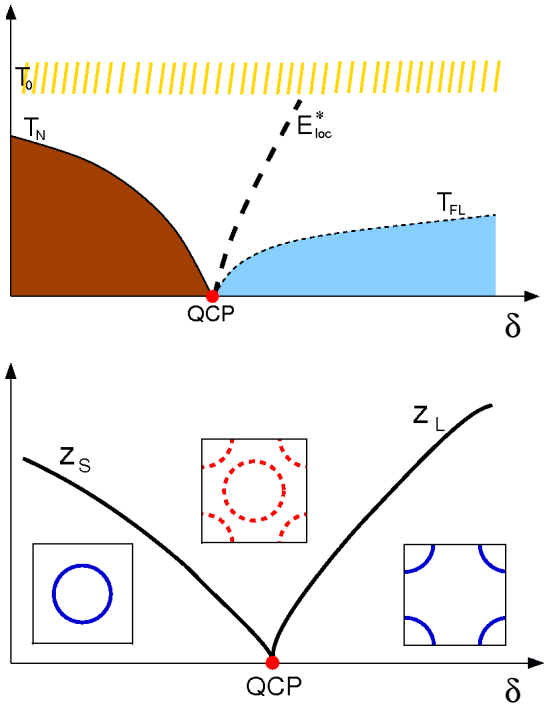}                  
\caption{(top) 
Local quantum criticality is characterized by the collapse of an energy scale,
$E_{\rm loc}^{\ast}$, at the continuous onset of AF order. Here $T_0$ is a crossover temperature scale 
marking the initial onset of Kondo screening as temperature is lowered,
and $\delta$ is the non-thermal tuning parameter representing the ratio
of the bare Kondo scale to the RKKY interaction. $T_{\rm N}$ is the N\'eel transition temperature,
and $T_{\rm FL}$ is the temperature scale for the heavy Fermi liquid state with a 
Kondo-entangled 
ground state. 
The QCP occurs at $\delta=\delta_c$.
(bottom)
Collapse
of the quasiparticle residues across the  local QCP.
 $z_{\rm L}$and $z_{\rm S}$ are quasiparticle residues for the small (left inset)
 and large (right inset) Fermi surfaces, respectively.
 At the QCP, the quasiparticles
are critical on both the small and large Fermi surfaces.
}
\label{localQCP}
\end{figure}

\section{Antiferromagnetic quantum critical points}\label{AFQCP}
Weak metallic antiferromagnets such as Cr are well-described in terms of a
spin-density-wave (SDW) order. The 
order parameter is the staggered
magnetization. Correspondingly, its transition at $T = 0$ to a paramagnetic
metal phase is described by an SDW QCP \cite{Hertz,Moriya,Millis}.
This
description falls into  the Landau framework, with the order parameter
differentiating between the phases and the fluctuations of the order parameter
describing the quantum criticality.

Microscopically, this has been studied in terms of a  one-band Hubbard
model \cite{Hertz}.  The SDW order is induced by the Coulomb repulsion, and the
effective field theory is a quantum Ginzburg-Landau action \cite{Hertz},
\begin{eqnarray}
{\cal S}
\!  = \!\!
\int d {\bf q}~
\! \!
\frac{1}{\beta}
\sum_{i\omega_n}
(r + c {\bf q}^2 + |\omega_n|/\Gamma_{\bf q})
\, {\bf \phi} ^2
+
\int \!\!~
\!\!\, u \, {\bf \phi} ^4
+ \dots 
\label{S-Hertz}
\end{eqnarray}
which describes the fluctuations of the order parameter, $\phi$, in both space
($\bf x$) and imaginary time ($\tau$).  This yields an effective dimensionality
$d+z$, where $d$ is the spatial dimension and $z$ is the dynamic exponent. The
wavevector ${\bf q}$ and Matsubara frequency $\omega_n$ are reciprocal to ${\bf
x}$ and $\tau$, respectively.

Theoretical studies on QCPs in AF heavy fermion metals have led to
beyond-Landau
QCPs, which feature inherently quantum modes  besides the
order-parameter fluctuations. While this notion extends to general contexts, for
heavy fermion metals the emphasis has been on the critical modes associated with
the destruction of the Kondo effect. Such critical modes are  in addition to the
fluctuations of the AF order parameter.

This Kondo destruction has provided insights into early puzzles on dynamical
scaling in quantum critical heavy fermion systems, and led to predictions for
the nature of Fermi surfaces  and energy scales near the QCP that have been
extensively observed in magnetotransport and  quantum oscillation experiments.

\subsection{Kondo destruction and beyond-Landau QCP}\label{KDQCP}
Heavy fermion metals are microscopically described in terms of the Kondo lattice
Hamiltonian, Eq.\,(\ref{kondo-lattice-model}).
Their traditional understanding invokes a ground state with a Kondo
singlet between the local moments and conduction electrons. As described
earlier, this Kondo entanglement in the ground  state gives rise to an
excitation spectrum with a Kondo resonance per local moment. The resulting heavy
Fermi liquid state has a large Fermi surface, which incorporates  $1+x$
electrons per unit cell. In this way, the heavy quasiparticles near the Fermi
energy can undergo an SDW order, thereby leading to an SDW QCP.

The dynamical spin correlations, however, may invalidate the very process that
gives rise to the Kondo singlet ground state and the associated Kondo
resonances. Such considerations have led to the notion of a local QCP,
characterized by the physics of Kondo destruction. The phenomenon of Kondo
destruction was already studied in the earlier renormalization-group
(RG)
study
of the charge analogue of the  dynamical RKKY vs Kondo 
competition \cite{Si96.1}, as well as its generalization to the spinful case 
\cite{SmithSi,Sengupta,Si.99}.
Motivated by experiments
on CeCu$_{5.9}$Au$_{0.1}$ \cite{Schroder}, that we discuss in
Sect.\,\ref{exp_probes} below, concrete theoretical formulations for the 
Kondo destruction  were advanced \cite{Si-Nature,Colemanetal,Si-prb.03}. 
The theory
implied the collapse of the extra energy scale $E_{\rm loc}^{\ast}$,
a jump of the
Fermi  surface across the QCP, and the critical nature of the quasiparticles 
on the whole Fermi surface
at
the QCP (cf.\ Fig.\,\ref{localQCP}). The Kondo destruction was later also
studied in a fermionic slave-particle representation of the Kondo lattice
Hamiltonian  \cite{Sen04.1,PaulPepinNorman.07}
and in dynamical mean field
theory \cite{Deleo.08}. Finally, the notion of non-Fermi liquid  quasiparticles
on the whole Fermi surface (as opposed to only on the hot spots) has also been
emphasized in a recent work based on a self-consistent  method
\cite{WolfleAbrahams_prb11}.

The extended dynamical-mean-field theory (EDMFT) 
\cite{SmithSi-edmft,Chitra.00,Si96.1}
provides a means to study the weakening of the Kondo effect by the dynamical
effects of  RKKY interactions.  The reduction of the Kondo singlet amplitude is
described through 
the decrease
of the energy scale $E_{\mathrm{loc}}^{\ast}$.

At the  local  QCP, the Kondo destruction ($E_{\mathrm{loc}}^{\ast}=0$) occurs at
$\delta_c$, as shown in  Fig.\,\ref{localQCP}. In this case, the  local spin
susceptibility is
\begin{eqnarray}
\chi({\bf q}, \omega ) =
\frac{1}{f({\bf q}) + A \,(-i\omega)^{\alpha} W(\omega/T)} \quad .
\label{QS:chi-qw-T}
\end{eqnarray}
This form was derived with the aid of an $\epsilon$-expansion
RG
 approach
 \cite{Si-Nature,Si-prb.03}. 
 The exponent $\alpha$ has been found to be
 close to 0.75 (ranging from 0.72 to 0.83) \cite{Grempel.03,Zhu.07,Glossop.07}.

The Kondo destruction  implies that, as the QCP is approached from the
paramagnetic side, the  quasi-particle residue $z_L \propto (b^{\ast})^2 \rightarrow
0$, where $b^{\ast}$ is the strength of the pole of the conduction-electron
self-energy $\Sigma({\bf k},\omega)$;
this pole characterizes the Kondo resonance. In turn, electron excitations have a
non-Fermi liquid form  on the 
entire
large Fermi surface.

\subsection{Dynamical Kondo effect}\label{DKE}
The breakdown of the large Fermi surface  implies that the Fermi surface will be
small on the  antiferromagnetically ordered  side. This is the ${\rm AF_S}$
phase, which will be further discussed in the  next section.
A  dynamical
Kondo effect will still generate  a mass enhancement in the ${\rm AF_S}$ phase
\cite{ZhuGrempelSi,Zhu.07,Glossop.07}.

\begin{figure}[t!]%
\includegraphics[width=0.9\linewidth]{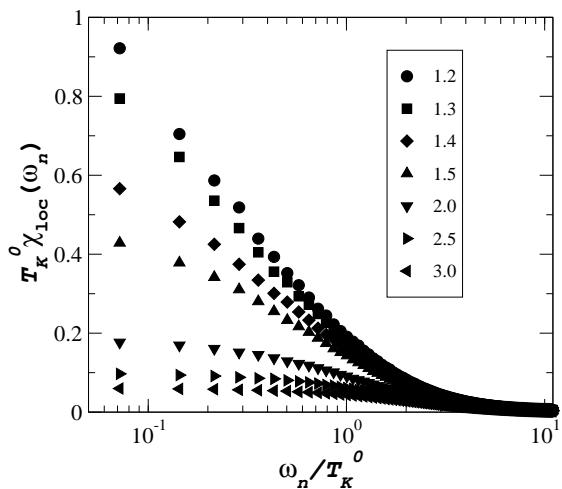}              
\caption{%
Dynamical Kondo effect inside the Kondo-destroyed AF phase.
Shown is the normalized local susceptibility vs. Matsubara frequency,
at a low temeperature ($T \approx 0.01T_K^0$).
The legend specifies $I/T_K^0$.
The gradual enhancement of the local dynamics as the 
QCP ($I_c/T_K^0=1.2$) is approached from the ordered side demonstrates the 
dynamical Kondo effect.
Figure from Ref.\,\cite{ZhuGrempelSi}.
}
    \label{dyn_Kondo}
\end{figure}

To consider this point further, the {\it static} amplitude of the Kondo singlet
vanishes in the ${\rm AF_S}$ phase. Thus, the electron distribution function in
the ground state displays a discontinuity at the small Fermi surfaces. By
contrast, the effective mass is a dynamical quantity, measuring the dispersion
of the Landau quasiparticles. The quasiparticle mass enhancement in the ${\rm
AF_S}$ phase arises due to the {\it dynamical} part of the Kondo-singlet
correlations, which persists in this phase. This has been explicitly
demonstrated in the calculated local dynamics in the ${\rm AF_S}$ region of the
Kondo lattice model \cite{ZhuGrempelSi,Zhu.07,Glossop.07},
as illustrated in Fig.\,\ref{dyn_Kondo}.

This dynamical Kondo effect ensures the continuity of the divergent effective
mass as the QCP is approached from both the AF ordered side and the paramagnetic
side.

\subsection{Experimental probes of QCPs}\label{exp_probes}
Numerous experimental techniques have been used to characterize quantum
critical materials. Both temperature-dependent measurements at different fixed
control parameter values and isothermal measurements as function of control
parameter have been instrumental in advancing the field \cite{Loe07.1}.
Temperature dependencies are frequently analysed by power law fits and then
presented in terms of colour-coded temperature--tuning parameter phase diagrams,
where the quantum critical region appears as a fan emerging from the QCP. Also
insightful are scaling analyses. They test in which temperature range and up to
which tuning-parameter distance from the QCP the data can be collapsed onto a
single universal curve, described by universal critical exponents. These
critical exponents are then compared to theoretical predictions.

A milestone was the scaling analysis of the dynamical spin susceptibility of
CeCu$_{\rm 6-x}$Au$_{\rm x}$ near the critical concentration 
${\rm x}_c = 0.1$, determined by
inelastic neutron scattering at different energy transfers $E$
\cite{Schroder}. The evidenced $E/T$ scaling and
the accompanying
anomalous
critical exponent are not expected in the order-parameter description of the SDW
QCP, and have been interpreted  in terms of Kondo destruction
\cite{Si-Nature,Colemanetal}.

Inelastic neutron scattering experiments on heavy fermion compounds are
demanding as the typically small magnetic moments require large single
crystalline samples and long data acquisition times.  Therefore, such experiments 
have only been used to study a limited number of quantum critical heavy fermion
metals; besides CeCu$_{\rm 6-x}$Au$_{\rm x}$, these include the earlier results
of  Aronson et al.\  on ${\rm UCu_{5-x}Pd_x}$  \cite{Aronson.95}  (which we will come back to
in Sect.\,\ref{Other}).
Nevertheless, further progress was made by exploring
materials with other experimental probes.

An important quantity is the Fermi surface. As it is expected to evolve
qualitatively differently across  an SDW and a local QCP
\cite{Colemanetal,Si-prb.03}, probes to detect its evolution are of great
interest. A simple measurement sensitive to changes of the Fermi surface is the
Hall coefficient $R_{\rm H}$. At $T = 0$, the Fermi surface is expected to
evolve continuously across 
an
SDW QCP but discontinuously, in the form of a jump,
across a local QCP. The former is due to the order parameter increasing
continuously as the ordered phase gets stabilized with increasing distance
$\delta$
from the critical tuning parameter value 
$\delta_{\rm
c}$. In the case of strong nesting, $R_{\rm H}$ may show a sizable (but
continuous) evolution within the ordered phase. Across a local QCP, on the other
hand, the Kondo entanglement is destroyed as the system orders. Thus, the $4f$
spins contribute to the Fermi surface only in the paramagnetic phase but drop
out of the Fermi surface in the ordered phase. This leads to a jump of the Fermi
surface. As the destruction of the Kondo entanglement is associated with an
temperature
scale $T^{\ast}$, that is in addition to the ordering temperature $T_{\rm
N}$ and collapses with it only at the QCP, experimental signatures of the Kondo
destruction are expected to occur away from $T_{\rm N}(\delta)$ at finite
temperatures. This is again in contrast to the SDW case, where $T_{\rm
N}(\delta)$ is the only 
temperature
scale and thus the only place in the phase
diagram where changes of the Fermi surface can be expected.

\begin{figure}[t!]                                                            
\includegraphics[width=\linewidth,height=0.743\linewidth]{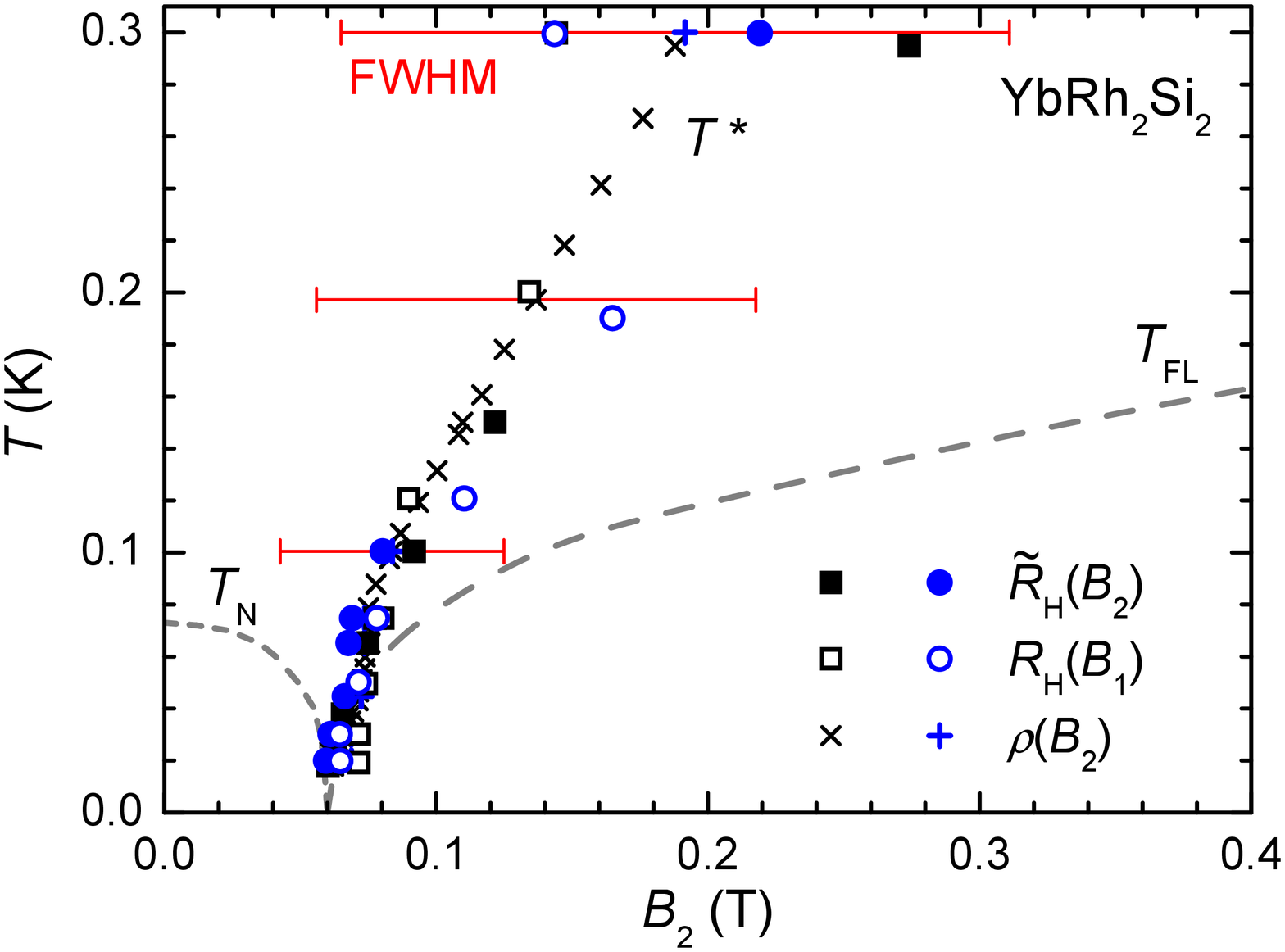}
\includegraphics[width=\linewidth,height=0.743\linewidth]{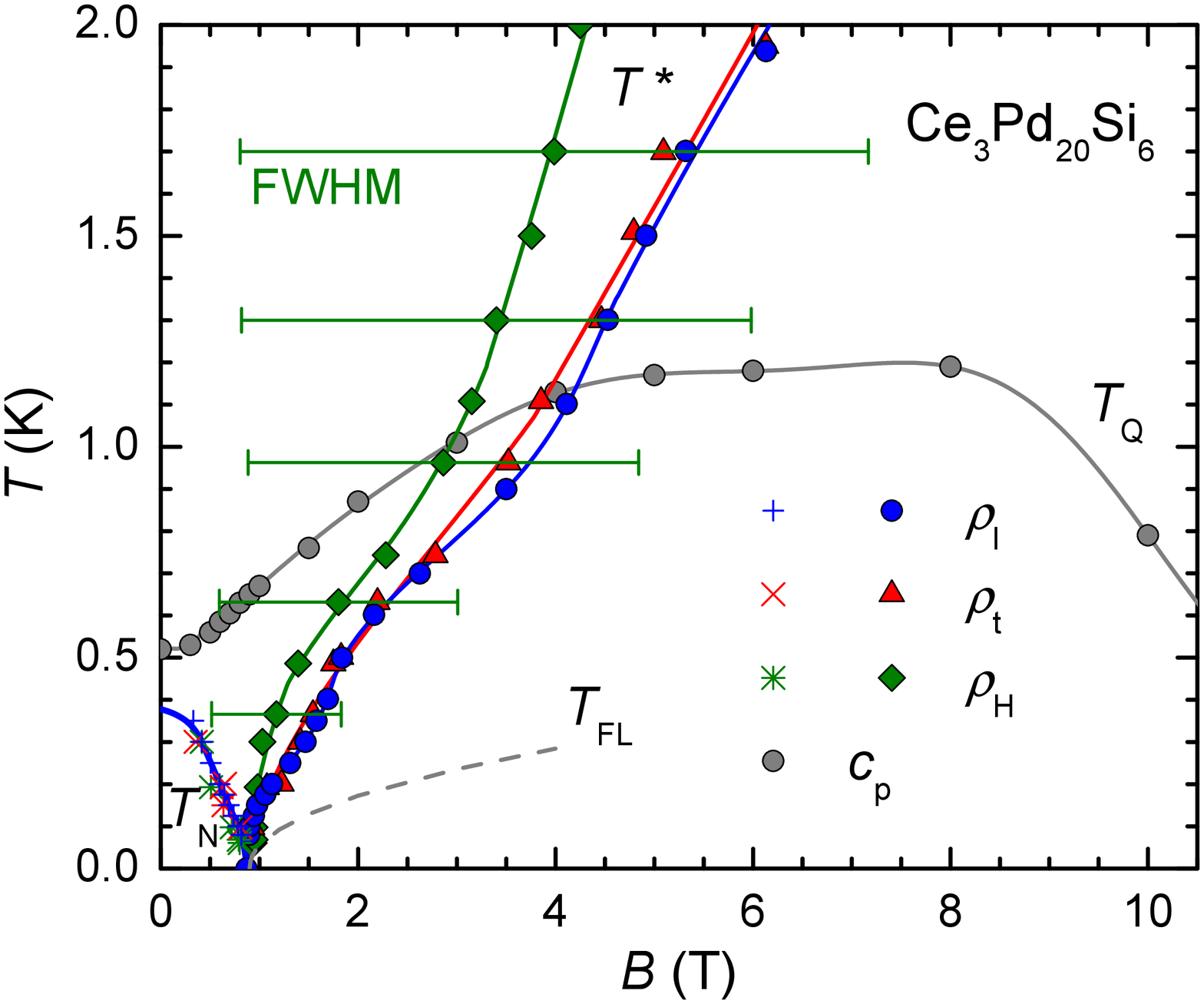}              
\caption{Temperature--magnetic field phase diagrams of YbRh$_2$Si$_2$ (top) and
Ce$_3$Pd$_{20}$Si$_6$ (bottom), showing the Kondo breakdown scale $T^{\ast}$, the
phase transition temperatures $T_{\rm N}$ and $T_{\rm Q}$, and the temperature
$T_{{\rm FL}}$ below which FL behaviour is recovered. The symbols indicate the
measurements from which the characteristic temperatures were extracted. For
YbRh$_2$Si$_2$: Differential Hall coefficient $\widetilde{R}_{\rm H}(B_2)$ in a
single-field experiment, Hall coefficient $R_{\rm H}(B_1)$ in a crossed-field
experiment, longitudinal resistivity $\rho(B_2)$; $B_1$ and $B_2$ are fields
$\parallel$ and $\perp$ to the $c$ axis. For Ce$_3$Pd$_{20}$Si$_6$: Longitudinal
resistivity $\rho_{\rm l}$, transverse resistivity $\rho_{\rm t}$, Hall resistivity
$\rho_{\rm H}$ and specific heat $c_{\rm p}$, all in single field experiments. FWHM
indicates the full width at half maximum of the crossover at $T^{\ast}$. In the
extrapolation to $T=0$, it collapses to zero for both materials. Figures adapted
from \cite{Fri10.2,Cus12.1}.}
\label{T-star}
\end{figure}

\begin{figure}[t!]                                                            
\includegraphics*[width=\linewidth]{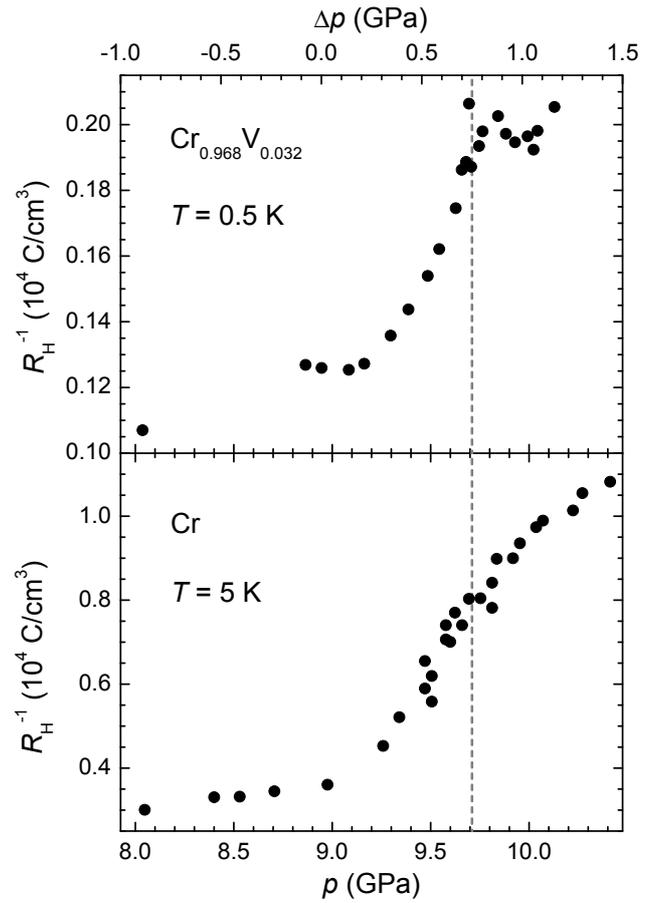}                
\caption{Evolution of the inverse Hall coefficient of Cr$_{\rm 1-x}$V$_{\rm x}$
across its pressure-driven QCP. For ${\rm x}=0.032$ (top), tuning is done both by
external and chemical pressure, for ${\rm x}=0$ (bottom) only external pressure was
applied. The continuous evolution of $R_{\rm H}^{-1}(p)$ confirms the absence of a
jump in the Hall coefficient across 
an
SDW QCP. Figures adapted from
\cite{Lee04.1,Jar10.1,Fri11.1}.}
\label{RH_Chromium}
\end{figure}

Of course, measurements of $R_{\rm H}$ can only be performed at finite
temperatures. Thus, it is important to detect both the position (tuning
parameter value at a given temperature) where a crossover in $R_{\rm H}$ is
observed and the sharpness of the crossover (e.g.\ in the form of the full width
at half maximum, FWHM). To distinguish between  the two QCP scenarios,
extrapolations of both quantities to $T = 0$ are needed. So far, this has been
reliably done for both YbRh$_2$Si$_2$ \cite{Pas04.1,Fri10.2} and
Ce$_3$Pd$_{20}$Si$_6$ \cite{Cus12.1}. In both systems, a crossover in $R_{\rm
H}$ was observed across a scale $T^{\ast}$, that coincides with $T_{\rm N}$ only
at the QCP. The FWHM of the crossover was shown to extrapolate to zero in the $T
= 0$ limit (Fig.\,\ref{T-star}), thus evidencing the presence of a local QCP in
both cases. 
(In Ce$_3$Pd$_{20}$Si$_6$, this occurs inside the ordered region of the phase diagram,
see Fig.\,\ref{T-star}. This is further discussed in the next section.)
By contrast, a combined hydrostatic and chemical pressure study of
the itinerant antiferromagnet Cr$_{\rm 1-x}$V$_{\rm x}$ revealed a continuous
evolution of $R_{\rm H}$ across the SDW QCP, with a width of the drop of $R_{\rm
H}^{-1}$ within the ordered phase below 9.7\,GPa that is essentially independent
of temperature (Fig.\,\ref{RH_Chromium}).

A second important tool to probe Fermi surface changes are de Haas--van
Alphen (dHvA) experiments. These have been used to study the Fermi surface
evolution of CeRhIn$_5$ as a function of pressure \cite{shishido2005}
at magnetic fields around 10 T, just above 
the upper critical field for the suppression of superconductivity $H_{c2}$.

In this material,  the pressure-induced AF to non-magnetic quantum phase
transition has been observed  to be second order for fields above 3 T 
\cite{park-nature06,Knebel.11}. The
second-order nature  of this transition is further evidenced by the tendency of
divergence near the critical pressure, $p_c$, for both   the cyclotron mass
$m^{\ast}$  \cite{shishido2005}
and the $T^2$-resistivity coefficient $A$  \cite{Knebel.08}.
Across such
an AF QCP, the
dHvA frequencies undergo a sharp
jump.
In the AF ordered state, at $p<p_c$, they are compatible
with the Fermi surface being small, while those in the paramagnetic state, at
$p>p_c$, are consistent with the Fermi surface being large. Such a sudden jump
of the Fermi surface across the QCP is again to be contrasted with the  smooth
evolution as indicated by the Hall-coefficient results across the SDW QCP in the
Cr:V system. 

We close this subsection by remarking on 
three aspects. 
First, as we discussed,
the dynamical Kondo effect leads to mass enhancement in the ${\rm AF_S}$ phase. 
Within the local QCP description, this dynamical Kondo effect underlies the
mass enhancement observed in the AF phases of YbRh$_{\rm 2}$Si$_{\rm 2}$,
CeCu$_{\rm 6-x}$Au$_{\rm x}$ and CeRhIn$_{\rm 5}$.  Beyond these,
recent experimental observations in the AF state of CeIn$_{\rm 3}$
at ambient pressure
appear to provide evidence for a dynamical Kondo 
effect as well \cite{Lizuka.12}.
CeCu$_{\rm 2}$Ge$_{\rm 2}$, in its AF state,  may also belong to ${\rm AF_S}$, 
and a similar
observation has also  been reported recently \cite{Bosse.12}.


\begin{figure*}[t!]
\centering\includegraphics*[width=0.75\linewidth]{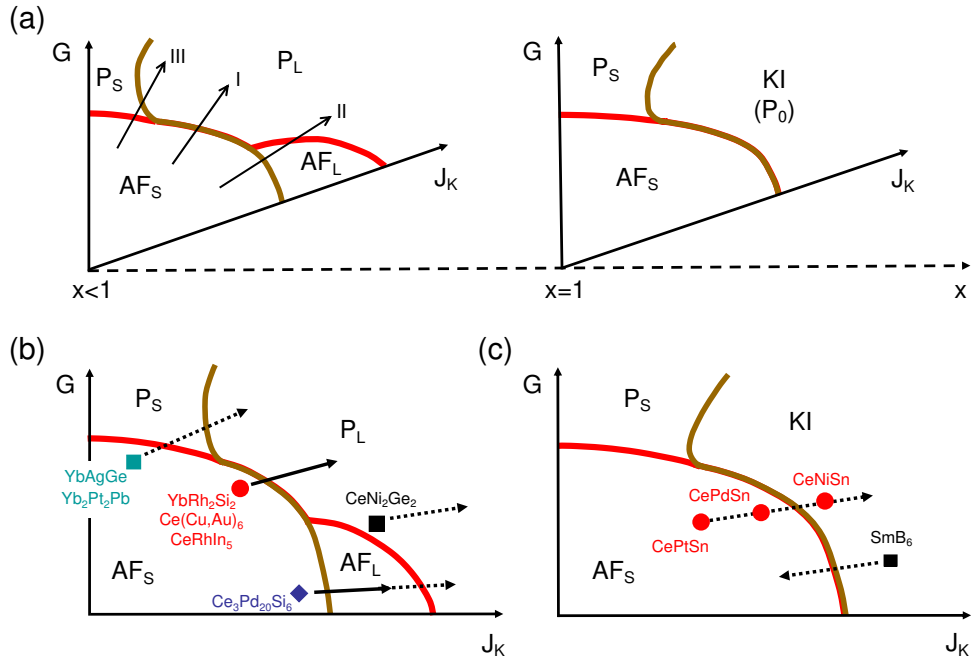}
\caption{(a) Global phase diagram for heavy fermion metals (left) and Kondo insulators (right), 
linked by the tuning parameter conduction band filling $x$. 
$G$ denotes the degree of quantum fluctuations of the local-moment magnetism,
and $J_K$ is the normalized Kondo coupling.  The phases are differentiated by being 
paramagnetic (${\rm P}$) or antiferromagnetic (${\rm AF}$), with a large Fermi surface
(${\rm L}$) which incorporates the $f$-electrons or a small 
Fermi
surface (${\rm S}$)
which does not. 
In the left panel, the three lines with arrow describe three trajectories 
of transitions from ${\rm AF_S}$ to ${\rm P_L}$. In the right panel, the Kondo insulator (KI)
phase, which is paramagnetic and has no Fermi surface (${\rm 0}$)
due to filled bands, replaces the ${\rm P_L}$ phase of the heavy fermion metal case;
correspondingly, there exists no ${\rm AF_L}$ phase.
Also shown are materials
in the global phase diagrams for heavy fermion metals (b) and Kondo insulators (c).
}\label{Global}
\end{figure*}


Second, the critical destruction of quasiparticles at the QCP has recently been probed 
by thermal transport measurements in YbRh$_2$Si$_2$, which observed a violation of the
Wiedemann-Franz law \cite{Pfau-nature12}.  STM spectroscopy is also poised to probe the nature 
of the single-electron excitations in the quantum critical regime \cite{Aynajian-nature12}.

Third,
recent inelastic neutron scattering measurements on YbRh$_2$Si$_2$
have finally observed the AF wavevector in the fluctuation  spectrum at $T>T_N$
\cite{Sto12.1}. This is consistent with the order being AF,  which has in the
past been inferred from the facts that magnetic field applied in any direction
reduces the ordering temperature and that the magnetic susceptibility is reduced
below the ordering temperature. Unlike the size of the Fermi surface, the dynamical spin
susceptibility {\it per se} does not unambiguously address the question of
whether $f$ electrons are localized or delocalized in the AF state. Still, the
incommensurate wavevector is stable up to quite high energy,  extending
to as large as 1.25~meV, which is cloe to the (bare) Kondo energy scale; 
this would not be expected for 
an
SDW order associated with the delocalized
$f$ electrons and appears to be more consistent with 
local-moment magnetism.
It would be instructive to see how the AF wavevector may connect
to the form of the RKKY interactions.

This neutron study opens the door to future measurements of the dynamical spin
susceptibility of YbRh$_2$Si$_2$ at low frequencies and temperatures that
may be able to address the nature of dynamical scaling in the quantum critical
regime.  In addition, at magnetic fields that correspond to the field-induced FL
regime (Fig.\,\ref{localQCP}, top), a resonance mode has been observed
\cite{Sto12.1}, which promises to shed light on the large-Fermi surface regime
of the phase diagram.


\section{Global phase diagram of antiferromagnetic heavy fermion metals}\label{GlobalHF}
The notion of Kondo destruction  not only provides a means to characterize novel
QCPs but also opens up the  possibility of new phases at zero temperature.

\subsection{Theoretical basis}
\label{gpd_theory}
A zero-temperature  phase diagram has been proposed for  the AF heavy fermion
metals \cite{Si.06,Si10.1,Col10.2}.
Figure~\ref{Global}(a), left panel,
shows the two-parameter phase diagram.
Here the horizontal axis is the Kondo coupling $J_K$, and the vertical axis $G$
marks the quantum fluctuations of the local-moment magnetism; the latter is
tuned by frustration and dimensionality.

The starting point 
of the theoretical basis
 is the
establishment of  AF$_{\rm S}$,
an antiferromagnetic phase 
with destruction of the Kondo effect and the concomitant small Fermi surface.
The demonstration of its stability was based on an analysis of the Kondo
lattice Hamiltonian in the $J_K \ll I \ll W$ limit. 
This limit allows an expansion 
around
 a new reference point,
corresponding to $J_K=0$, where the ordered local-moment system is decoupled
from the non-interacting conduction electrons.
With respect to this reference point, the effect of $J_K$ can be systematically
studied using a renormalization-group method.
For the Ising
case, the AF-ordered phase of the local-moment magnetism has a spin gap; as a
result, $J_K$ is irrelevant implying the stability of the ${\rm AF_S}$ phase
\cite{Si.06}.

For the case with full spin-rotational invariance, corresponding to an SU(2)
symmetry, the analysis is more involved. Using a quantum non-linear sigma model
(QNL$\sigma$M) representation, and with the aid of a mixed
gapless-boson-and-fermion 
RG
method, the Kondo
coupling is 
shown to be 
exactly marginal, leading to the stability of the ${\rm
AF_S}$  phase \cite{Yamamoto.07}.

A sufficiently large $J_K$ gives rise to the Kondo entanglement,
resulting in the standard heavy Fermi liquid state -- a  paramagnetic phase 
with a large Fermi surface.
Therefore, 
 ${\rm AF_S}$ and ${\rm P_L}$ 
 represent two anchoring
phases
of the phase
diagram in Fig.\,\ref{Global}(a), left panel. The distinction between these two phases are 
beyond-Landau: in addition to the presence/absence of AF order, the two are also differentiated 
by the absence/presence of Kondo entanglement. 

The global phase diagram features three types of transition sequences as the system
goes from  
the ${\rm AF_S}$ phase to the ${\rm P_L}$ phase.
The type I transition
goes directly from  
${\rm AF_S}$ to ${\rm P_L}$,
while
the type II sequence  corresponds to transitions from ${\rm AF_S}$ through ${\rm
AF_L}$ to ${\rm P_L}$. In other words, they differ 
in terms of 
whether the collapse
of the Kondo effect occurs at the  paramagnetic-to-AF QCP, or whether it is
delayed until inside the ordered part of the phase diagram. This distinction was
inferred from
EDMFT studies \cite{Si-Nature,Si-prb.03}. The ${\rm AF_L}$ phase 
corresponds to the SDW order of the heavy fermion quasiparticles
of the ${\rm P_L}$ phase, and Kondo
destruction  occurs at the boundary between ${\rm AF_L}$ and ${\rm AF_S}$
phases. Note that, again, the dynamical Kondo effect  in the ${\rm AF_S}$ phase
ensures the continuity of the quasiparticle  mass across its boundary with the
${\rm AF_L}$ phase.

The type III sequence from  ${\rm AF_S}$ to ${\rm P_L}$ passes through
${\rm P_S}$, a paramagnetic phase with Kondo destruction and hence small Fermi
surface. The existence of the ${\rm P_S}$ phase can be seen by noting that
increasing $G$ of the local-moment  magnetism alone (along the vertical axis)
suppresses AF order. In the most typical cases, this leads to spin-Peierls
order, which contains a spin gap. This spin gap renders a small $J_K$ coupling
to be irrelevant in the RG sense, thereby yielding the stability of a ${\rm
P_S}$ phase.  The competition between the ${\rm P_L}$ and ${\rm P_S}$ phase 
has been analyzed through the effect of the Berry phase in the  QNL$\sigma$M
representation \cite{Goswami.11}.

\subsection{Materials basis}

In Sect.\,\ref{exp_probes}, we have discussed the evidence for a quantum phase
transition along the trajectory I (${\rm AF_S}$ directly to ${\rm P_L}$) in pure
YbRh$_2$Si$_2$ as a function of the magnetic field, in CeCu$_{\rm 6-x}$Au$_{\rm x}$ as a
function of
the substitution level 
${\rm x}$ or pressure, and in CeRhIn$_5$ as a function of
pressure at magnetic fields just above $H_{c2}$. These three cases are
illustrated in Fig.\,\ref{Global}(b).

Systematic investigations have also been performed on substituted and
pressurized YbRh$_2$Si$_2$  \cite{Friedemann09,Cus10.1,Tok09.1,GegenwartXX.1}.
The resulting substitution level--magnetic field phase diagram is
consistent with the profile of the global phase diagram in
Fig.\,\ref{Global}(b). In particular, compounds with Ni- and slight
Ir-substitution for Rh also undergo a type I transition (${\rm AF_S}$ to ${\rm
P_L}$). For the case of Co-substitution for Rh as well as for the pure
YbRh$_2$Si$_2$ under sufficient pressure, the transition appears to be divided
into two stages, as along the type II trajectory with the Kondo destruction
occurring inside the AF order. By contrast, for substitutions with Ge, or Ir
concentrations beyond 2.5\%, the transition seems to be of type III, a
trajectory with the Kondo destruction outside the ordered phase. Direct
measurements of the Kondo destruction in substituted
and pressurized YbRh$_2$Si$_2$ remain, however, to be be carried out.

To explore the global phase diagram, it is essential to 
broaden both the materials basis and the
microscopic parameter variety.
In particular, it is instructive to consider materials that lie 
in
different ranges of the parameter $G$, the degree of quantum 
fluctuations in the local-moment magnetism.
A recent investigation of the cubic 
compound Ce$_3$Pd$_{20}$Si$_6$  \cite{Cus12.1} 
probed the lower part of the global phase diagram, where $G$ 
is small (dimensionality high) 
because of the absence of 
 crystal anisotropy. A magnetic field-tuned QCP was observed to be accompanied
 by Kondo destruction. 
However, the phase above the critial field (cf.\ Fig.\,\ref{T-star}) still possess some kind
of order.
While both the type of order and the nature of the transition to the 
paramagnetic phase 
need to be explored in future experiments, these results provide strong evidence 
for a type II trajectory.

In another cubic material, CeIn$_3$, dHvA experiments revealed a strong
enhancement of the effective mass at a critical field within the ordered portion
of the temperature--magnetic field phase diagram \cite{Seb09.1}. Thus, one may
speculate that a type II trajectory is followed also by this material. In order
to further test the global phase diagram, the role of dimensionality might also
be studied more systematically, e.g.\ by the variation of the thickness of
thin films of heavy fermion compounds. That such tuning might, in fact, be
technically feasible was demonstrated by recent investigations on molecular beam
epitaxy grown superlattices \cite{Shi10.1,Miz11.1}.

Geometrical frustration is expected to be a means to enhance $G$ and  reach the
upper portion of the global phase diagram. Earlier studies considered the
hexagonal YbAgGe \cite{Bud05.2}.
More recently, Yb$_2$Pt$_2$Pb was investigated \cite{Kim-prl.13}, in which
the Yb moments are located on a quasi-two-dimensional Shastry-Sutherland
lattice.
In both cases, there exists an intermediate magnetic field regime between an AF
order and a paramagnetic Fermi liquid regime, as illustrated in 
Fig.\,\ref{Global}(b). Other materials in this category might be  the fcc YbPtBi
\cite{Mun.13} and CePdAl with a distorted Kagome lattice \cite{Fritsch12}.

Clearly, this is a subject that will see major developments in the near future, both theoretically
and experimentally.  These studies may ultimately lead us towards a classification of universality
classes in quantum critical heavy fermions.


\section{From local moment to mixed valent regime}\label{MV}

The Kondo lattice Hamiltonian is only valid when the underlying $f$-electron
orbital is half-filled, with $n_f=1$ per unit cell. Most of the heavy fermion
metals displaying AF QCPs, such as YbRh$_2$Si$_2$ and CeCu$_{6-x}$Au$_x$, show a
very large mass enhancement and a
small Kondo temperature on the order of 10\,K,
and should therefore be in the  Kondo limit.
This 
is
also 
expected to 
be the case
for CeRhIn$_5$, for which the Kondo temperature is
similarly
about 10\,K \cite{Hegger.00}.
However, some materials do
not appear to be in this limit. For instance, 
in the quantum
critical superconductor $\beta$-YbAlB$_4$,
the mass enhancement is modest, the Kondo temperature is rather high (on the order of 200\,K),
and the Yb valence
was shown to be 2.75 
(distinctly smaller than the integer valence 3)
by hard X-ray
photoemission spectroscopy \cite{Oka10.1}.
 While the occurrence of mixed valence 
is rare in heavy fermion metals, it is more common in Kondo insulators. 
In fact, these latter were originally referred to 
as mixed valent semiconductors \cite{Wac94.1}. For the prototypal Kondo insulator SmB$_6$
\cite{Men69.1},
mixed valence was detected early on \cite{Van65.1}. 
Thus, it is interesting to look at the generalization 
of the Kondo lattice Hamiltonian to
the mixed valent regime, where $n_f \neq 1$. 

\subsection{Periodic Anderson Model}
The underlying microscopic Hamiltonian 
in the mixed valent regime
is
the 
periodic Anderson model
\begin{eqnarray}
H_{\rm{PAM}} =
& & \sum_{\vec{k}\sigma}
\epsilon_{\vec{k}}c_{\vec{k}\sigma}^{\dagger}c_{\vec{k}\sigma} 
+
 \sum_{\vec{k}\sigma}
(V_{\vec{k}}f_{\vec{k}\sigma}^{\dagger}c_{\vec{k}\sigma}+ \rm{H.C.})
\nonumber \\
& & 
+
\epsilon_f
\sum_{i} n_{f,i}
+
U\sum_{i}
n_{f,i\uparrow}n_{f,i\downarrow}
\quad ,
\label{PAM}
\end{eqnarray}
which contains both spin and charge (valence) degrees of freedom
of the $f$ electrons.
Here, $n_{f,i\sigma} = f_{i\sigma}^{\dagger}f_{i\sigma}$
and $n_{f,i} = \sum_{\sigma} n_{f,i\sigma}$.

Away from the local-moment limit, the inter-orbital Coulomb interaction 
\begin{equation}
U_{fc} \sum_{i}
n_{f,i}
n_{c,i}\quad ,
\end{equation}
that was introduced early on \cite{FalicovKimball69,Varma76},
also has to be considered.
 In the mixed valent regime, it influences the charge Kondo effect and the 
corresponding mixed valent non-Fermi liquid behavior
\cite{SiKotliar93,Perakis93,SiJPCM96}.

\subsection{Gaussian fixed point: generalization of the SDW QCP to mixed valence}
In the Kondo lattice case, assuming a finite Kondo energy scale at the QCP led
to the description of the SDW QCP (Sect.\,\ref{KDQCP}). This description has
been generalized to the mixed valence setting \cite{Watanabe-prl10}.
Microscopically, for
$U_{fc}$ below a critical value $U_c$,
 there is a valence crossover 
($n_f$ varies continuously with $\epsilon_f$), 
whereas
for $U_{fc} > U_c$ 
there is a 
first 
order valence transition
($n_f$ shows a discontinuous decrease at some $\epsilon_f$). 
The quantum-critical end point is a Gaussian fixed
point, representing the generalization of the SDW QCP to the co-existing 
spin and charge sector.
Indeed, the effective field theory of Ref.\, \cite{Watanabe-prl10}
has the form of the Hertz action,
Eq.\,(\ref{S-Hertz}), with the damping $\Gamma_{\bf q} \propto q$ 
corresponding to a dynamic exponent $z=3$.
The QCP retains the generic feature
of a Gaussian fixed point, in that $\omega/T$ scaling is violated; presumably
$H/T$ scaling is correspondingly violated as well.

Like in the SDW QCP, there will be no jump of the Fermi surface across 
such a QCP.
Indeed, in order to realize a jump in the Fermi surface (and the Hall coefficient)
in this picture, 
the transition is taken to be first order \cite{Wat10.1}.
Alternatively, it was suggested that a jump in the Fermi surface can arise 
in a continuous transition due to folding of the large Fermi surface by the AF
order \cite{Watanabe-jpcm12}.
This, however, appears incorrect: as the example of Cr:V (Fig.\,\ref{RH_Chromium})
demonstrated, the Hall coefficient must be continuous across such a QCP.

\begin{figure}[h]%
\includegraphics[width=0.6\linewidth,angle=270]{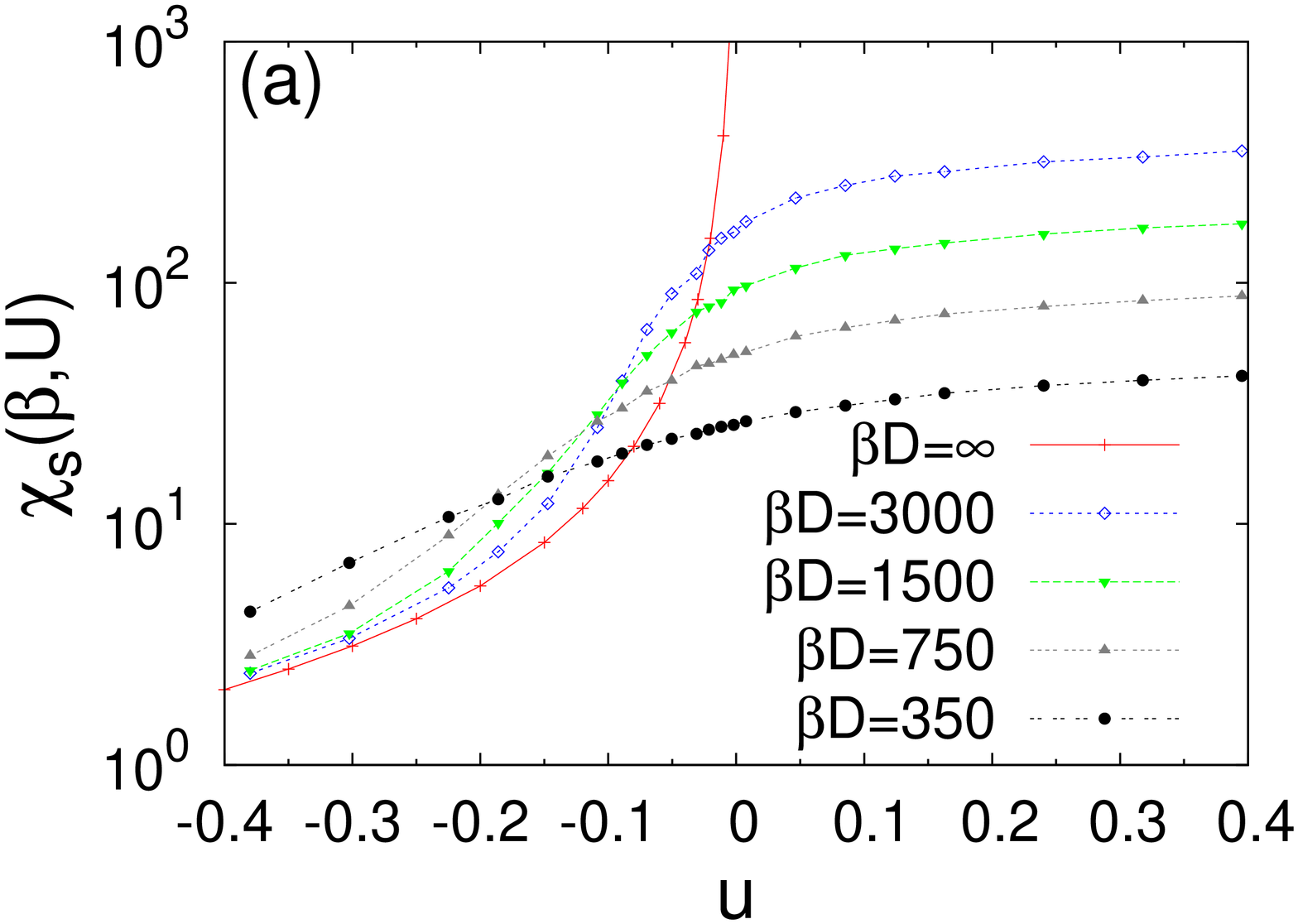}
\includegraphics[width=0.6\linewidth,angle=270]{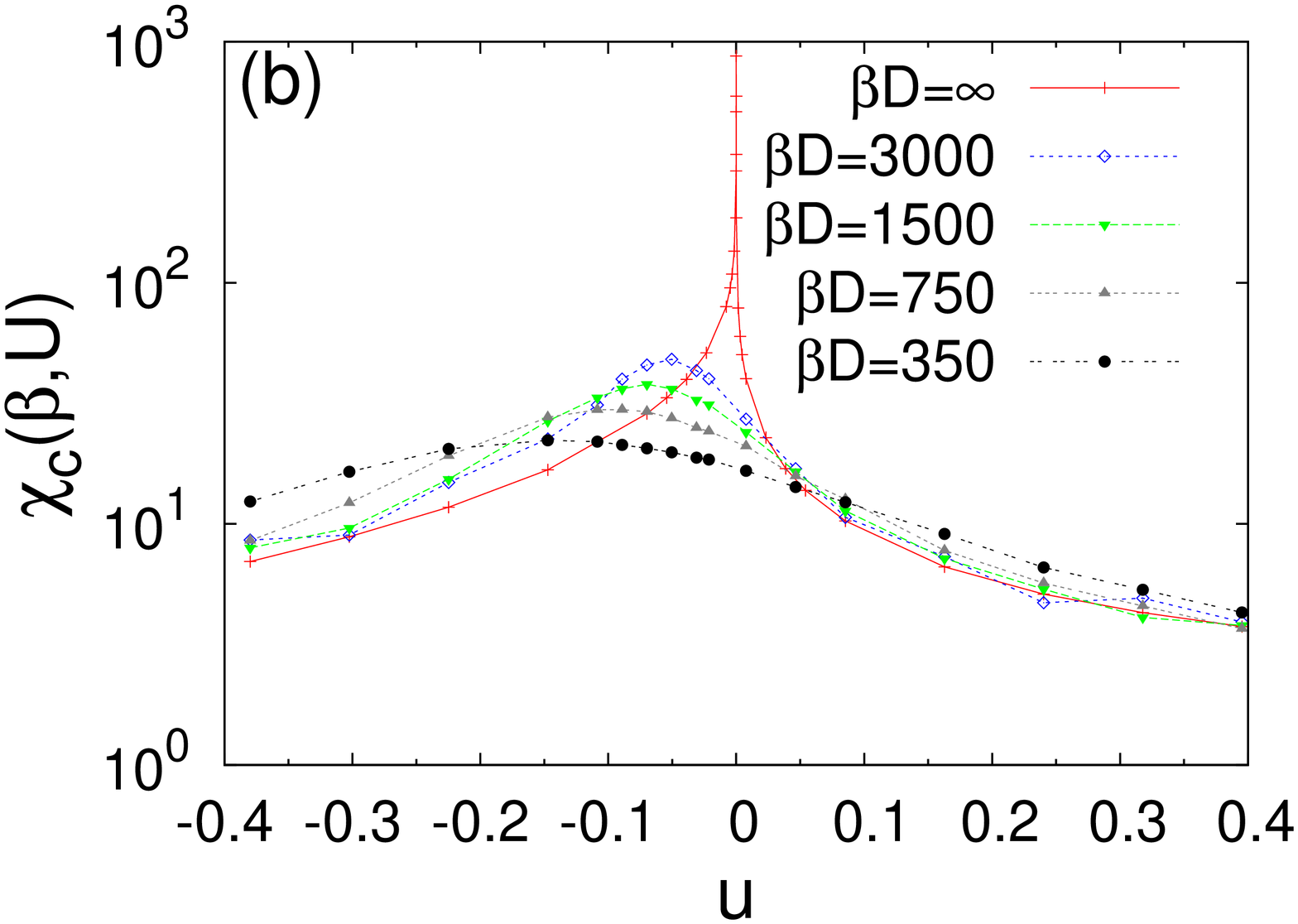}   
\includegraphics[width=0.6\linewidth,angle=270]{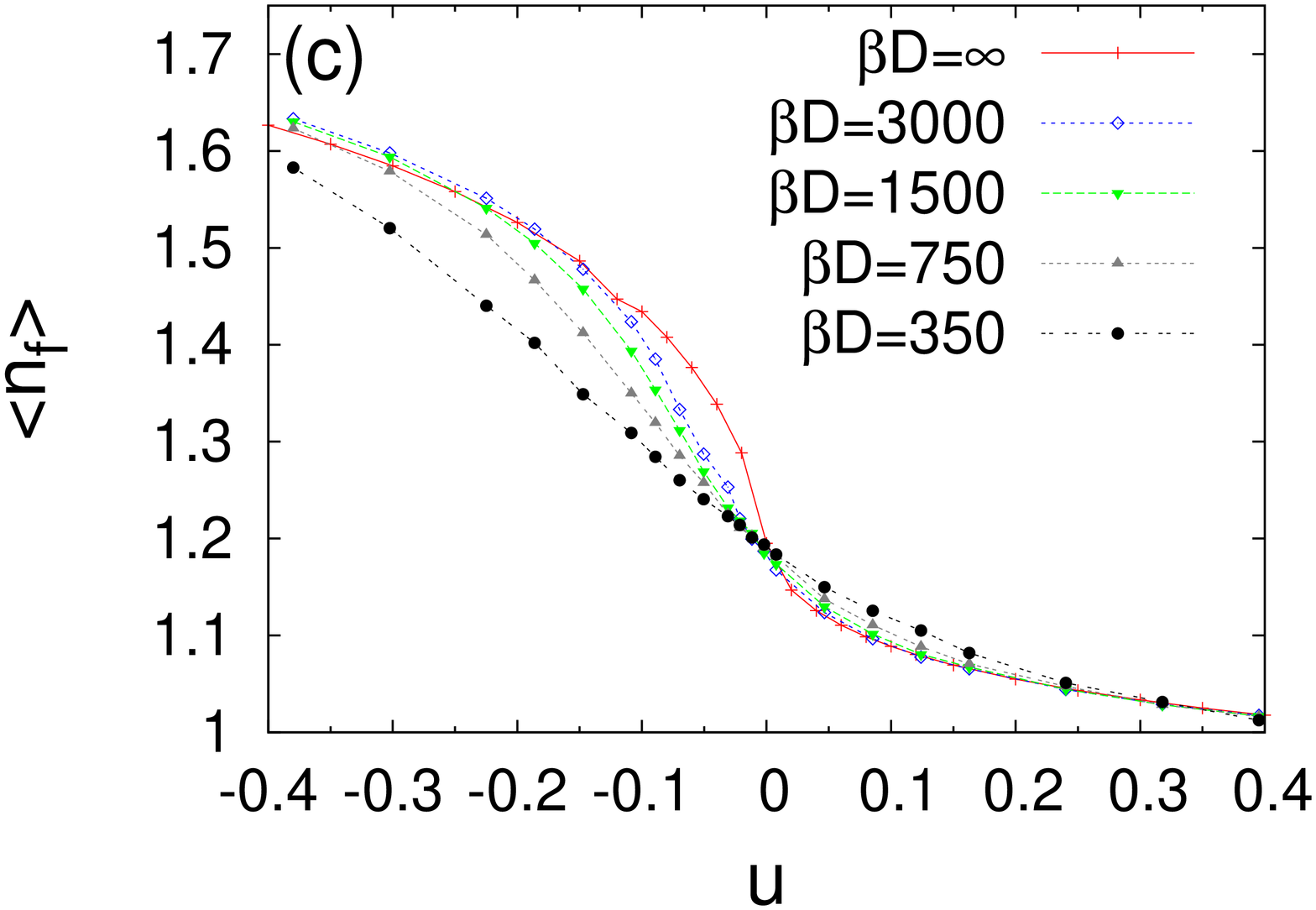}                 
\caption{%
Kondo destruction in a mixed valence model.
(a) Local static spin susceptibility $\chi_{s}$
vs. the tuning parameter \ $u=U/U_c\!-\!1$ at various temperatures $T=1/\beta$,
measured in units of the half-bandwidth $D$;
its divergence on approaching the QCP  ($u=0$) at $T=0$ signifies the Kondo destruction,
with an associated Kondo energy scale in the spin sector going 
to zero as the QCP is approached from 
the Kondo-screened side.
(b) Local static charge susceptibility $\chi_{c}$ is also divergent at the QCP.
(c) Occupancy $\langle n_f \rangle$ vs the control parameter;
at the QCP, $\langle n_f \rangle \neq 1$. Figures from Ref.\,\cite{Pix12.1}.
}
    \label{Kondo-destruction-mv}
\end{figure}

\begin{figure*}[t!]    
\includegraphics*[width=0.9\linewidth]{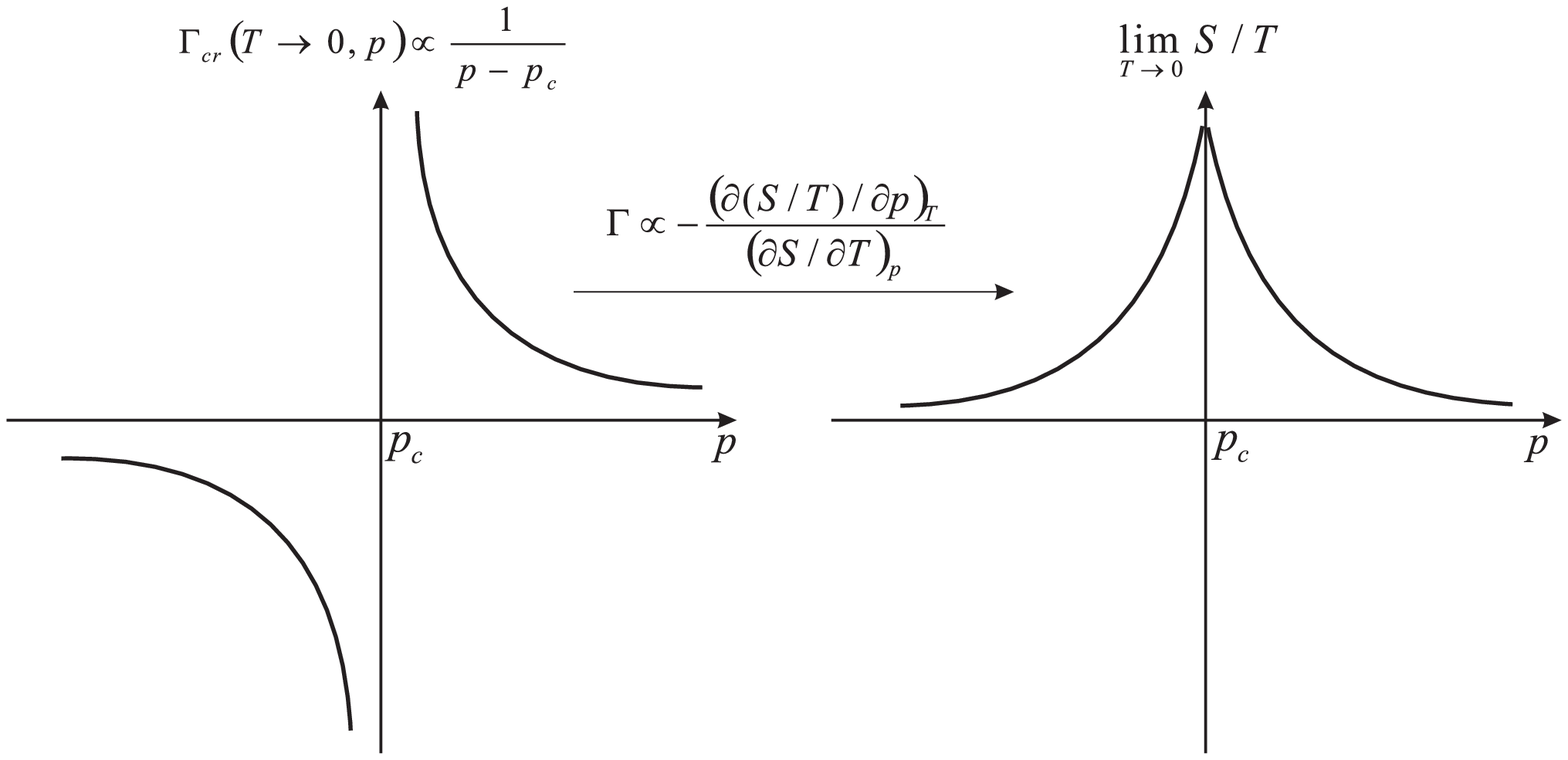}              
\caption[]{
Entropy accumulation near QCP and divergence of the Gr\"uneisen 
ratio,  both of which follow
from the scaling form of the free energy \cite{Zhu03.1}.
Figure from Ref.\,\cite{Wu-jpcs10}.
}
\label{entropy}
\end{figure*}

\subsection{Interacting fixed point: generalization of the
Kondo destruction to mixed valence}

To search for an interacting fixed point,
 it is important to address
whether the Kondo destruction effect can still take place 
under the mixed valence
condition.
This is a delicate issue. 
In the Kondo lattice case, Kondo destruction amounts to the localization of $f$
electrons. This is physically transparent, because localization can readily
arise for a commensurate filling of an electronic orbital (one $f$ electron per
site). At mixed valence, the $f$ orbital has a fractional, generally
incommensurate, per-site occupancy; there is no mechanism known for electron
localization in such cases.

This issue has recently been studied in a simplified setting, the pseudo-gapped
particle-hole-asymmetric Anderson impurity model. It was shown that Kondo 
destruction occurs even at mixed valence \cite{Pix12.1}, see
Fig.\,\ref{Kondo-destruction-mv}.  The physics in the spin sector is very
similar to that for the local-moment case $n_f = 1$, in that an energy scale
collapses as the QCP is approached from the Kondo-screened side. At the QCP, the
charge fluctuations are singular along with the spin fluctuations. 
(This is in contrast to the particle-hole symmetric local-moment case $n_f=1$,
where the singular charge fluctuations are absent.)
In line with
the interacting nature of the fixed point, both the charge and  spin responses
display a magnetic field over temperature ($H/T$) and frequency over
temperature  ($\omega/T$) scaling. When the results are generalized to the
lattice case, the Kondo scale going to zero would imply that the Fermi
surface jumps.

Recent measurements on the mixed valent heavy fermion compound
$\beta$-YbAlB$_4$ have  shown that the static magnetization exhibits $H/T$
scaling  \cite{Nakatsuji-science11,Nak08.1}, providing evidence for the
interacting nature of an underlying QCP.  The above 
theoretical
study  in a simplified model
provides 
an existence proof
of a local QCP at mixed valence with this kind
of scaling behavior.  If the QCP displays universality, such scaling behavior 
would also be expected at other mixed valent QCPs with Kondo destruction,
irrespective of the microscopic details that give rise to valence fluctuations.

The theoretical results also suggest that,  by restraining our further
discussion to the Kondo case $n_f = 1$, we do not lose generality.


\section{Global phase diagram of Kondo insulators}\label{GlobalKI}

In the case of half-filling, i.e.\ $x=1$, the competition between the Kondo and RKKY
interactions continues to operate. 
This case presents an opportunity
to study a variety of correlated insulator states.

\subsection{Global phase diagram}
As discussed in Sect.\,\ref{KondoLattice}, when Kondo entanglement takes place,
the ground state for the case of
half-filling is a paramagnetic insulator, the Kondo insulator.
Thus, in the
global phase diagram, instead of a heavy fermion metal state with 
a
large Fermi
surface (P$_{\rm L}$) we now have a state with no Fermi surface at all (P$_0$).
Both the AF$_{\rm S}$ and P$_{\rm S}$ phases are
expected to remain stable in the case of half-filling \cite{Yam10.1}
(Fig.\,\ref{Global}(a), right panel).

There is 
evidence
 that transitions between these phases can indeed
be realized experimentally. While tuning through a QCP appears to be
accomplished in a number of materials as function of magnetic field
\cite{Coo99.1,Jai00.1} or doping \cite{Bus98.1,Kal00.1,Sle05.1,Slu12.1}, we
focus here on the conceptually simpler situation of tuning by external or
chemical pressure. As an example, SmB$_6$ is tuned by pressure from the Kondo-insulating phase P$_0$ to an AF phase \cite{Bar05.1}. The dashed
arrow in Fig.\,\ref{Global}(c) suggests that this transition is to the phase
AF$_{\rm S}$. The explicit experimental demonstration that the AF phase of
SmB$_6$ has a small Fermi 
surface
is, however, still 
missing.

While most Kondo insulators are cubic, there are also a few representatives with
lower 
crystalline
symmetry, the most prominent one being orthorhombic CeNiSn. The
isostructural and isoelectronic compounds CePtSn and CePdSn both order
antiferromagnetically at low temperatures \cite{Kas88.1,Mal89.1,Adr94.1}.
Transitions between the AF phase and the Kondo-insulating phase have been
investigated both by chemical and by hydrostatic pressure tuning. The former has
been realized by isoelectronic substitutions, i.e.\ in Ce(Pt$_{1-x}$Ni$_x$)Sn
\cite{Sak92.1,Adr96.1,Kal00.1} or Ce(Pd$_{1-x}$Ni$_x$)Sn \cite{Kas91.1}, and by
hydrogenation \cite{San07.1}. Tuning by hydrostatic pressure appears more
challenging. The magnetic order in CePdSn
was shown to be suppressed only by
hydrostatic pressure above 60~kbar \cite{Iga93.1}. A pressure of 20~kbar applied
to CePtSn reduces the N\'eel temperature only slightly \cite{Als11.1}. The
ensemble of these experiments is represented schematically in
Fig.\,\ref{Global}(c) by the dashed arrow through CePtSn, CePdSn and CeNiSn. As
for SmB$_6$, an explicit demonstration of a small Fermi surface AF$_{\rm S}$ in
the antiferromagnetically ordered state still needs to be done. The lower
symmetry of CeNiSn compared to SmB$_6$ is represented in Fig.\,\ref{Global}(c)
by placing CeNiSn and its relatives at a higher $G$ value than pressure tuned
SmB$_6$.

It would be instructive to consider other Kondo insulator systems in the 
context of the global phase diagram. For instance, 
CeRu$_2$Al$_{10}$ and CeFe$_2$Al$_{10}$
\cite{Strydom09}
might fit into this framework
as well.

\subsection{Correlated topological insulators and their interplay with Kondo coherence and magnetim}
In recent years, spin-orbit coupling in band insulators
has been extensively
studied as giving rise to topological insulators \cite{HasanKane10,QiZhang11}.
These bulk insulators  have gapless helical surface states. In Kondo insulators,
topological insulating states have been studied either through  the spin-orbit
coupling as manifested in the hybridization matrix \cite{Dze10.1}
or in conjunction with
the spin-orbit coupling
of the  conduction electrons \cite{Feng.12}. 
As such, Kondo
insulators present an opportune  system to study the interplay between
topological and Kondo/magnetic effects, 
and even
unconventional
superconductivity near such boundaries.  
Recent measurements of several groups
in SmB$_6$
have provided tentative evidence for the surface states of a topological
insulator 
\cite{Wolgast12,Botimer12}.
The evolution of such surface states under pressure
[cf.\
Fig.\,\ref{Global}(c)]
represents an intriguing  issue for future studies.

From the perspective of topological states \cite{Feng.12},
it is instructive to 
note that the  5$d$ electrons of CePtSn will contain a large spin-orbit coupling,
which may influence the topological nature of its expected insulating state 
at high pressure.


\section{QCPs in other degrees of freedom}\label{Other}

\subsection{Weak ferromagnets and ferromagnetic heavy fermion metals}
Based on the field theory of Hertz for the ferromagnetic case,
Eq.\,(\ref{S-Hertz}), the coupling to the fermions near the Fermi surface 
has been shown to yield non-analyticity in the coupling constants
\cite{belitz-rmp05}.
This is believed to turn the $T = 0$ transition first order,
creating a ``fish-tail''
structure in the
temperature--pressure--magnetic field phase diagram
\cite{Uhlarz04}.

In Kondo lattice systems, the effect of local moments must be considered.
For the ferromagnetic Kondo lattice, Ref.\, \cite{Yamamoto-pnas10}
studied the ordered phase using the
QNL$\sigma$M basis (generalizing the study for the AF ordered phase
mentioned in Sect.\,\ref{gpd_theory}).
Inside the ferromagnetic phase, the Fermi surface is small.
The existence of this ${\rm F_S}$ phase raises the prospect that 
the $T=0$ ferromagnetic  to paramagnetic  transition may also come in different 
varieties, in a way that is analogous to the global phase diagram of the AF 
heavy fermion metals described earlier. In particular, 
this raises the intriguing possibility for a QCP in ferromagnetic 
heavy fermion metals.

The early dHvA measurements \cite{King.91}
in the ferromagnetic compound CeRu$_2$Ge$_2$ (with 
a Curie temperature $T_c = 8$\,K)
can be interpreted in terms of the ${\rm F_S}$ phase. 
Other ferromagnetic heavy fermion compounds of interest include 
the Re-doped URu$_2$Si$_2$ \cite{Butch-prl09},
CeRu$_2$Al$_2$B \cite{Baumbach-prb12},
and YbNi$_3$Al$_9$ \cite{Miyazaki-prb12}.
Recent experiments have provided evidence for a ferromagnetic QCP,
both in YbNi$_4$P$_2$ \cite{Krellner-njp11}
and in the iron pnictide
Ce(Ru$_{\rm 1-x}$Fe$_{\rm x}$)PO
\cite{Kitagawa-prl12}.

\subsection{Other degrees of freedom}

Quadrupolar order has
also been discussed in the context of quantum criticality
in a number of heavy fermion compounds. In the filled skutterudite
PrOs$_4$Sb$_{12}$,
antiferroquadrupolar order 
is believed to characterize a high-field ordered phase (HFOP), that is adjacent to a
superconducting phase  \cite{Ho03.1}.
Thus, the possibility arises that the superconductivity in
PrOs$_4$Sb$_{12}$ is 
driven
by quantum critical fluctuations of quadrupolar
degrees of freedom. In UPd$_3$ antiferroquadrupolar order can be
weakened
 by
small amounts of Np. At 5\% doping, where the order is just suppressed,
unconventional 
properties are
observed, 
implicating a potential
QCP
\cite{Wal07.1}.

UCu$_{5-x}$Pd$_{x}$ is the earliest system in which a dynamical  spin 
susceptibility
with an anomalous critical exponent and $E/T$ scaling was discovered
\cite{Aronson.95}. There have been theoretical studies which attributed 
this behavior to the effects of disorder, including a distribution 
of Kondo temperature scales \cite{Miranda-rpp05}
or a Griffiths phase \cite{CastroNeto-prb00}. However, it could also
result from an interacting spin-glass QCP with
Kondo destruction \cite{Si.06},
that appears to 
occur
 in Sc$_{1-x}$U$_{x}$Pd$_3$.


\section{Novel phases near quantum critical points}\label{Emergent}

QCPs
tend to nucleate
 new phases,
 as illustrated by the observation of 
 superconductivity near heavy fermion QCPs \cite{Mathur98}.
From a macroscopic perspective, general scaling considerations show that 
entropy is maximized near a QCP \cite{Zhu03.1,Wu-jpcs10}.
This is illustrated in Fig.\,\ref{entropy}.
Such an accumulation of entropy has been explicitly demonstrated across the
field-induced metamagnetic QCP 
in 
Sr$_3$Ru$_2$O$_7$ \cite{Rost.09}.
An enhanced entropy suggests that the system is particularly prone 
to developing new order. 
In the global phase diagram, 
we may consider $AF_L$ and $P_S$ as new phases that emerge in the vicinity of QCPs.
Another instructive case is provided by CeAgSb$_2$, which 
possesses
ferromagnetic order
below about 10\,K at ambient pressure. A pressure of about 3 GPa reduces
the Curie temperature to about 4\,K. Before the ferromagnetic order is 
completely suppressed, however, it succumbs to an AF state
\cite{Sidorov-prb03}.
In other words, the AF state emerges at the border of
a ferromagnetic order.

Of course, the most striking example
is superconductivity  developing near  AF QCPs. We have discussed the
strong evidence that the AF quantum phase  transition in CeRhIn$_5$ is described
by a local QCP. A superconducting state  of a very large  $T_c$ (when
measured against the Kondo temperature) in CeRhIn$_5$ \cite{Hegger.00}
occurs in the vicinity  of its AF QCP \cite{park-nature06,Knebel.11}.
Hence, this material provides a striking
example 
for superconductivity that is 
driven by Kondo destruction, developing from a normal state with fluctuating Fermi surfaces
and 
critical quasiparticles.

In the case of CeCu$_2$Si$_2$, there is evidence that the asymptotic 
low-energy quantum critical behavior is of the SDW type. Nonetheless,
the exchange energy gain
is
much larger 
than the condensation energy,
which has been interpreted
\cite{Stockert-natphys11,Stockert-jpsj12}
in terms of a
Kondo-destruction 
energy
scale
that
is nonzero but relatively small compared to the Kondo scale. This suggests 
that
Kondo destruction physics is part of the dynamics that drives 
superconductivity in CeCu$_2$Si$_2$.

\section{Summary and outlook}\label{Summary}

We close with some perspectives and outlook.

The study of Kondo destruction and local quantum criticality has 
broad implications. The notion
that quasiparticles break down over 
the
 entire Fermi surface represents a drastic
departure from the traditional 
SDW description of an antiferromagnetic
quantum critical point. 
In the
SDW case, quasiparticles 
disintegrate only near hot spots, the parts of the Fermi surface that are connected 
by the antiferromagnetic wavevector. We have discussed how the 
non-Fermi liquid behavior 
intertwines
with the additional modes of quantum criticality, 
and how
this quantum critical behavior
has been experimentally probed through observations that pertain to dynamical scaling, 
Fermi surface jump, extra energy scale and mass divergence.

The notion of Kondo destruction has also provided a means to characterize new quantum phases 
in antiferromagnetic heavy fermion metals. The resulting global phase diagram contains novel phases 
near the quantum critical points, that are not expected when the phases are only distinguished in 
terms of 
spontaneously broken 
symmetries
as in the Landau framework. 
Experimental evidence is growing for the phases that are differentiated by both Fermi surfaces 
and magnetic order parameter. Furthermore, the global phase diagram is fueling the recent 
efforts to extend the basis of heavy fermion materials with varying dimensionality 
or frustration-enhanced quantum fluctuations.

These insights on the normal state will undoubtedly impact on our understanding 
of superconductivity in heavy fermion metals, which tends to occur near quantum critical points.  
The spin-density-wave quantum critical excitations provide the basis for the spin-fluctuation theory 
of heavy fermion superconductivity, which assumes that a large Fermi surface is fully formed 
and quasiparticles near the hot spots are coupled to the critical spin fluctuations. 
Given the considerable evidence for local quantum criticality, it will be important to theoretically 
address how superconducting pairing 
arises out of such a drastic non-Fermi liquid state.

The insights gained on these issues, regarding non-Fermi liquid behavior and unconventional
superconductivity,  will likely be relevant to the understanding of other strongly 
correlated electron systems. Several additional aspects of heavy fermion physics also
promise to have implications in broader contexts.
The development of local quantum criticality has taken place through the Kondo destruction physics
in heavy fermion metals. As an interacting critical theory involving inherently quantum modes, 
it promises to connect with the physics of quantum phase transitions in other contexts.
It has already influenced the development of quantum criticality in insulating magnets.
More recently, 
tantalizing connections have been drawn between local quantum 
criticality in Kondo lattice systems and
criticality emerging
from holographic models in weakly curved anti-de Sitter (AdS) spacetime through the
string/field theory duality
\cite{Iqbal-prd10,Yam10.1,Sachdev-prl10};
``semi-local" critical points with closely related 
form of  correlation  functions have been constructed 
from the gravity side
\cite{Iqbal11,Faulkner-jhep11}.
New insights are likely to come from exploring this connection further.

Recent efforts, both theoretical and experimental,
were devoted to
spin-orbit coupling and topological states in heavy fermion systems. 
When spin-orbit couplings are explicitly taken into account,
the tuning among the Kondo insulator and antiferromagnetic states we discussed 
for the Kondo insulator systems
would  
be generalized to 
the transformation
among the topological 
and Kondo coherent/magnetic states. 
It is therefore likely that
Kondo insulator
systems 
will serve as a promising setting 
to study correlation effects 
on topological insulating states.

\begin{acknowledgement}
We would like to thank many colleagues for collaborations and discussions,
including E. Abrahams, P. Coleman, J. Custers, J. Dai, S. Friedemann,
P. Gegenwart, C. Geibel, P. Goswami, K. Ingersent,
S. Kirchner, C. Krellner, H. Liu,
A. H. Nevidomskyy, E. Nica, J. Pixley, A. J. Schofield, 
F. Steglich, O. Stockert, A. M. Strydom,  S. Wirth, J. Wu, 
S. Yamamoto, 
R. Yu,
H. Q. Yuan, J.-X. Zhu, and L. Zhu. 
This work has been supported in part by the
NSF Grant No. DMR-1006985, 
the Robert A. Welch Foundation Grant No. C-1411,
and the
European Research Council under the European Community's Seventh
Framework Programme (FP7/2007-2013)/ERC grant agreement no. 227378.
\end{acknowledgement}




\end{document}